\documentclass[pra,aps,showpacs,groupedaddress,superscriptaddress,twocolumn]{revtex4-1}

\usepackage[utf8x]{inputenc}
\usepackage{color}
\usepackage{bbm} 

\usepackage{amsfonts,amsmath,amssymb,stmaryrd}

\usepackage{graphicx}
\usepackage{subfigure}  
\usepackage{bbm} 
\usepackage{hyperref}
\usepackage{epsfig}
\usepackage{mathrsfs}
\usepackage{verbatim}
\usepackage{centernot}
\usepackage{ulem}

\renewcommand{\l}{\left(}
\renewcommand{\r}{\right)}

\newcommand{\gs}{\text{gs}}

\newcommand{\bra}[1]{\langle#1|}
\newcommand{\ket}[1]{|#1\rangle}

\renewcommand{\H}{\hat{\mathcal{H}}}

\renewcommand{\a}{\hat{a}}

\newcommand{\ad}{\hat{a}^\dagger}

\newcommand{\G}{\hat{\Gamma}}

\newcommand{\hc}{\text{h.c.}}
\newcommand{\MF}{\text{MF}}

\newcommand{\I}{\text{I}}
\newcommand{\p}{\text{p}}
\newcommand{\f}{\text{F}}
\newcommand{\s}{\text{S}}
\renewcommand{\sf}{\text{MIX}}

\newcommand{\F}{\hat{F}}

\newcommand{\ph}{\text{ph}}
\newcommand{\IB}{\text{IB}}
\newcommand{\B}{\text{B}}

\usepackage{array}

\def\smallint{\begingroup\textstyle \int\endgroup}

\usepackage{cancel,ifthen}
\newcommand{\cmnt}[2][NoInPuT]{\ifthenelse{\equal{#1}{NoInPuT}}{}{{\color{red}\sout{#1}}} {\color{blue} #2}}

\usepackage{bm}	
\renewcommand{\vec}[1]{\bm{#1}}

\begin{document}
\normalem	

\title{Renormalization group approach to the Fr\"ohlich polaron model: \\ application to impurity-BEC problem}

\author{F. Grusdt}
\affiliation{Department of Physics and Research Center OPTIMAS, University of Kaiserslautern, Germany}
\affiliation{Graduate School Materials Science in Mainz, Gottlieb-Daimler-Strasse 47, 67663 Kaiserslautern, Germany}
\affiliation{Department of Physics, Harvard University, Cambridge, Massachusetts 02138, USA}

\author{Y. E. Shchadilova}
\affiliation{Russian Quantum Center, Skolkovo 143025, Russia}
\affiliation{Department of Physics, Harvard University, Cambridge, Massachusetts 02138, USA}

\author{A. N. Rubtsov}
\affiliation{Department of Physics, Moscow State University, 119991 Moscow, Russia}
\affiliation{Russian Quantum Center, Skolkovo 143025, Russia}

\author{E. Demler}
\affiliation{Department of Physics, Harvard University, Cambridge, Massachusetts 02138, USA}


\date{\today}

\begin{abstract}
We develop a renormalization group approach for analyzing Fr\"ohlich polarons and apply it to a problem of impurity atoms immersed in a Bose-Einstein condensate of ultra cold atoms. Polaron energies obtained by our method are in excellent agreement with recent diagrammatic Monte Carlo calculations for a wide range of interaction strengths. We calculate the effective mass of polarons and find a smooth crossover from weak to strong coupling  regimes. Possible experimental tests of our results in current experiments with ultra cold atoms are discussed.
\end{abstract}

\maketitle

\section{Introduction}

A general class of fundamental problems in physics can be described as an impurity particle interacting with a quantum reservoir. This includes Anderson's orthogonality catastrophe \cite{Anderson1967}, the Kondo effect \cite{Kondo1964}, lattice polarons in semiconductors, magnetic polarons in strongly correlated electron systems and the spin-boson model \cite{Leggett1987}. The most interesting systems in this category can not be understood using a simple perturbative analysis or even self-consistent mean-field (MF) approximations. For example, formation of a Kondo singlet between a spinful impurity and a Fermi sea is a result of multiple scattering processes \cite{Anderson1961} and its description requires either a renormalization group (RG) approach \cite{Wilson1975} or an exact solution \cite{Andrei1980,Wiegmann1983}, or introduction of slave-particles \cite{Read1983}. Another important example is a localization delocalization transition in a spin bath model, arising due to "interactions" between spin flip events mediated by the bath \cite{Leggett1987}. 

While the list of theoretically understood non-perturbative phenomena in quantum impurity problems is impressive, it is essentially limited to one dimensional models and localized impurities. Problems that involve mobile impurities in higher dimensions are mostly considered using quantum Monte Carlo (MC) methods \cite{Prokofev1998,Gull2011,Anders2011}. Much less progress has been achieved in the development of  efficient approximate schemes. For example a question of orthogonality catastrophe for a mobile impurity interacting with a quantum degenerate gas of fermions remains a subject of active research \cite{Rosch1999,Knap2012}.

\begin{figure}[b!]
\centering
\epsfig{file=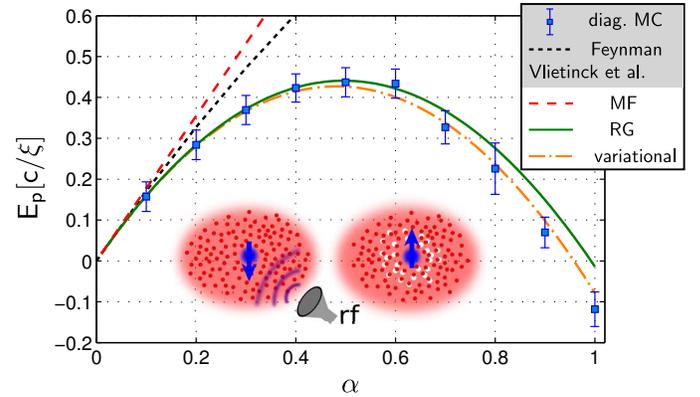, width=0.5\textwidth}
\caption{By applying a rf-pulse to flip a non-interacting (left inset) into an interacting impurity state (right inset) a Bose polaron can be created in a BEC. From the corresponding rf-spectrum the polaron groundstate energy can be obtained. In the main plot we compare polaronic contributions to the energy $E_\p$ (as defined in Eq.\eqref{eq:EpDef}) predicted by different models, as a function of the coupling strength $\alpha$. Our results (RG) are compared to Gaussian variational calculations \cite{Shchadilova2014}, MC calculations by Vlietinck et al. \cite{Vlietnick2014}, Feynman variational calculations by Tempere et al. \cite{Tempere2009} and MF theory. We used the standard regularization scheme to cancel the leading power-law divergence of $E_\p$. However, to enable comparison with the MC data, we did \emph{not} regularize the UV log-divergence reported in this paper. Hence the result is sensitive to the UV cutoff chosen for the numerics, and we used the same value $\Lambda_0=2000 / \xi$ as in \cite{Vlietnick2014}. Other parameters are $M/m=0.263158$ and $P=0$.}
\label{fig:FIG1}
\end{figure}

Recent experimental progress in the field of ultracold atoms brought new interest in the study of impurity problems. 
Feshbach resonances made it possible to realize both Fermi \cite{Schirotzek2009,Nascimbene2009,Koschorreck2012,kohstall2012metastability,Zhang2012,Massignan_review} 
and Bose polarons \cite{Catani2012,Fukuhara2013} with tunable interactions between the impurity and host atoms. Detailed information
about Fermi polarons was obtained using a rich toolbox available in these experiments. Radio frequency (rf) spectroscopy was used to measure the polaron binding energy and to observe the transition between the 
polaronic and molecular states \cite{Schirotzek2009}. The effective mass of Fermi polarons was studied using measurements of 
collective oscillations in a parabolic confining potential \cite{Nascimbene2009}. Polarons in a Bose-Einstein condensate (BEC) received 
less experimental attention so far although polaronic effects have been observed in nonequilibrium dynamics
of impurities in 1d systems \cite{Palzer2009,Catani2012,Fukuhara2013}.

The goal of this paper is two-fold. Our first goal is to present a new theoretical technique for analyzing
a common class of polaron problems, the so-called Fr\"ohlich polarons. We develop a unified approach that can describe 
polarons all the way from weak to strong couplings. Our second goal is to apply this method to the problem of impurity atoms immersed in a BEC. We focus on calculating the polaron binding energy and effective mass, both of which can 
be measured experimentally. Considering a wide range of atomic mixtures with tunable interactions \cite{Chin2010} and very different mass 
ratios available in current experiments \cite{Egorov2013,Spethmann2012,Pilch2009,Lercher2011,mccarron2011dual,Catani2008,Wu2012a,Park2012,Schreck2001,truscott2001observation,Shin2008,Bartenstein2005,Roati2002,Ferlaino2006,Ferlaino2006err,Inouye2004,Scelle2013,Hadzibabic2002,Stan2004,Schuster2012,schmid2010dynamics}, 
we expect that many of our predictions can be tested in the near future. In particular we discuss that currently available technology should make it possible to realize intermediate coupling polarons.

Previously the problem of an impurity atom in a superfluid Bose gas has been studied theoretically using the weak coupling
MF ansatz \cite{BeiBing2009,Shashi2014RF}, the strong coupling approximation \cite{Cucchietti2006,Sacha2006,Bruderer2007,Bruderer2008,Casteels2011}, the variational Feynman path integral approach \cite{Tempere2009}, and the numerical diagrammatic MC simulations \cite{Vlietnick2014}. 
The four methods predicted sufficiently different polaron binding energies in the regimes of intermediate and strong interactions, see FIG.\ref{fig:FIG1}. While the MC result can be considered as the most reliable of them, the physical insight gained from this approach is limited.
Our new method builds upon earlier analytical approaches by considering fluctuations on top of the MF state and including correlations between different modes using the RG approach. We verify the accuracy of this method by demonstrating excellent agreement with the MC results \cite{Vlietnick2014} at zero momentum and for intermediate interaction strengths. 

Our method provides new insight into polaron states at intermediate and strong coupling by showing the importance of entanglement between phonon modes at different energies. A related perspective on this entanglement was presented in Ref. \cite{Shchadilova2014}, which developed a variational Gaussian wavefunction for Fr\"ohlich polarons. Throughout the paper we will compare our RG results to the results computed with this variational correlated Gaussian wavefunction approach. In particular, we use our method to calculate the effective mass of polarons, which is a subject of special interest for many physical applications and remains an area of much controversy. At the end, we also comment on extensions of our approach to non-equilibrium problems.

This paper is organized as follows. We introduce the Fr\"ohlich Hamiltonian in Section \ref{sec:Model} and discuss how it can be used to describe mobile impurity atoms interacting with the phonons of a BEC. In Section \ref{sec:RevMF} we review MF results for the Fr\"ohlich polaron and formulate a Hamiltonian that describes fluctuations on top of the MF state. We discuss the solution of this Hamiltonian using an RG approach in Section \ref{sec:RGana}. In Section \ref{sec:regularization} we investigate cutoff dependencies of the polaron energy and describe how they should be properly regularized. Section \ref{sec:Results} provides a detailed discussion of our results for both the polaron binding energy and the effective mass. We show that the effective mass should be a much better probe of beyond MF aspects of the system. In Section\ref{Supp:ExperimentalConsiderations} we discuss possible experimental realizations and challenges, before closing with an outlook in Section \ref{sec:outlook}.

\section{Fr\"ohlich Hamiltonian}
\label{sec:Model}
The Fr\"ohlich Hamiltonian represents a generic class of models in which a single quantum mechanical particle interacts with the phonon reservoir of the host system. In particular it can describe the interaction of an impurity atom with the Bogoliubov modes of a BEC \cite{Cucchietti2006,Bruderer2007,Tempere2009}. In this section we review this model in both its original form and following the Lee-Low-Pines (LLP) transformation into the impurity reference frame. The goal of the LLP transformation is to use conservation of the total momentum to eliminate the impurity degrees of freedom, at the cost of introducing interactions between phonon modes.   

Our starting point is the Fr\"ohlich Hamiltonian describing the interaction between an impurity atom and phonon modes of the BEC ($\hbar=1$)
\begin{eqnarray}
\H_{\rm FROL} &=& \H_{\rm PHON} + \H_{\rm IMP} +\H_{\rm INT},
\nonumber\\
\H_{\rm PHON} &=&  \int_k d^dk ~ \omega_k \ad_{\vec{k}} \a_{\vec{k}},
\nonumber\\
\H_{\rm IMP} &=& \frac{P^2}{2M},
\nonumber\\
\H_{\rm INT} &=& \int_{|k|<\Lambda_0} d^d k~ V_k (\ad_{\vec{k}} + \a_{-\vec{k}}) e^{i \vec{k} \cdot \vec{R}}.
\label{H_frolich}
\end{eqnarray}
Here $M$ denotes the impurity mass and $m$ the mass of the host bosons, $\a_{\vec{k}}$ is the annihilation operator of the Bogoliubov phonon excitation in a BEC with momentum $\vec{k}$, $\vec{P}$ and $\vec{R}$ are momentum and position operators of the impurity atom, $d$ is the dimensionality of the system and $\Lambda_0$ is a high momentum cutoff needed for regularization. 
The dispersion of phonon modes of the BEC and their interaction with the impurity atom are given by the standard Bogoliubov expressions \cite{Tempere2009}
\begin{eqnarray}
\omega_k &=& c k \sqrt{1+\frac{1}{2} \xi^2 k^2},
\nonumber\\
V_k &=& \sqrt{n_0} (2 \pi)^{-3/2} g_\IB \l \frac{(\xi k)^2}{2 + (\xi k)^2} \r^{1/4},
\end{eqnarray} 
with $n_0$ being the BEC density and $\xi = \l 2 m g_\text{BB} n_0 \r^{-1/2}$ the healing (or coherence) length and $c=\l g_\text{BB} n_0 / m \r^{1/2}$ the speed of sound of the condensate. Here $g_{\rm IB}$ denotes the interaction strength between the impurity atom with the bosons, which in the lowest order Born approximation is given by $g_{\rm IB}=2\pi a_{\rm IB}/m_{\rm red}$, where $a_{\rm IB}$ is the scattering length and $m_{\rm red}^{-1}=M^{-1}+m^{-1}$ is the reduced mass of a pair consisting of impurity and bosonic host atoms. Similarly, $g_\text{BB}$ is the boson-boson interaction strength. The analysis of the UV divergent terms in the polaron energy will require us to consider a more accurate cutoff dependent relation between $g_{\rm IB}$ and the scattering length $a_{\rm IB}$ (see Eq. (\ref{eq:LSE}) below).  

When we calculate the energy of the impurity atom in the BEC we need to consider the full expression $E_{\rm IMP}= E_{\rm IB}^0 + E_{\rm B}$, where $E_{\rm IB}^0 = g_{\rm IB} n_0$ is the MF interaction energy of the impurity with bosons from the condensate, and $E_\B = \langle \H_\text{FROL} \rangle_\text{gs}$ is the groundstate energy of the Fr\"ohlich Hamiltonian. From now one we will call $E_{\rm IB}^0$ the impurity-condensate interaction energy and $E_{\rm B}$ the polaron binding energy. As we discuss below only  $E_{\rm IMP}$ is physically meaningful and can be expressed in a universal cutoff independent way using the scattering length $a_{\rm IB}$. Precise conditions under which one can
use the Fr\"ohlich model to describe the impurity BEC interaction, and parameters of the model for specific cold atoms mixtures are discussed in Sec. \ref{Supp:ExperimentalConsiderations}. We point out that the Fr\"ohlich type Hamiltonians \eqref{H_frolich} are relevant for many systems besides BEC-impurity polarons. Its original and most common use is in the context of electrons coupled to crystal lattice fluctuations in solid state systems \cite{Froehlich1954}. Another important application area is for studying doped quantum magnets, in which electrons and holes are strongly coupled to magnetic fluctuations. Motivated by this generality of the model \eqref{H_frolich} we will analyze it for a broader range of parameters than may be relevant for the current experiments with ultra cold atoms.

The Hamiltonian \eqref{H_frolich} describes a translationally invariant system. It is convenient
to perform the Lee-Low-Pines (LLP) transformation \cite{Lee1953} that separates the system into decoupled sectors of conserved total momentum,
 \begin{eqnarray}
 \hat{U} &=& e^{i \hat{S}} \hspace{1cm} \hat{S}= \vec{R} \cdot \int d^d k ~ \vec{k} \ad_k \a_k
 \label{LLP transformation}
\\
\H_{\rm LLP}&=& \hat{U}^\dagger \H_\text{FROL} \hat{U} =  \frac{1}{2 M} \Bigl( \vec{P} - \int d^d k~\vec{k} \ad_k \a_k \Bigr)^2  +
\nonumber \\
& & + \int d^d k ~ \left[  \omega_k \ad_{\vec{k}} \a_{\vec{k}} + V_k \l \ad_{\vec{k}} + \a_{-\vec{k}} \r  \right]. 
\label{H_LLP} 
\end{eqnarray}
The transformed Hamiltonian \eqref{H_LLP} does no longer contain the impurity position operator $\vec{R}$. Thus $\vec{P}$ in equation 
(\ref{H_LLP}) is a conserved net momentum of the system and can be treated as a $\mathbb{C}$-number (rather than an operator). 
Alternatively, the transformation \eqref{LLP transformation} is commonly described as going into the impurity frame, since the term describing 
boson scattering on the impurity in \eqref{H_LLP} is obtained from the corresponding term in \eqref{H_frolich} by setting $\vec{R}=0$.
The Hamiltonian (\ref{H_LLP}) has only phonon degrees of freedom but they now interact with each other. This can be understood physically as a phonon-phonon interaction, mediated by an exchange of momentum with the impurity atom. This impurity-induced interaction between phonons in \eqref{H_frolich} is proportional to $1/M$. Thus in our subsequent analysis of the polaron properties, which is based on the LLP transformed Fr\"ohlich Hamiltonian, we will consider $1/M$ as controlling the interaction strength.

\section{Review of the Mean Field Approximation}
\label{sec:RevMF}
In this section we briefly review the MF approach to the polaron problem, which provides an accurate description of the system in the weak coupling regime. We discuss how one should regularize the MF interaction energy, which is UV divergent for $d \geq 2$. To set the stage for subsequent beyond MF analysis of the polaron problem, we derive the Hamiltonian that describes fluctuations around the MF state.

The MF approach to calculating the ground state properties of \eqref{H_LLP} is to consider a variational wavefunction in which all phonons are taken to be in a coherent state \cite{Lee1953}. The MF variational wavefunction reads
 \begin{eqnarray}
| \psi_{\rm MF} \rangle = e^{\int d^3 \vec{k} ~  \alpha^\MF_{\vec{k}} \ad_{\vec{k}} - \hc } \ket{0} = \prod_{\vec{k}} \ket{\alpha_{\vec{k}}}.
\label{Psi_MF}
 \end{eqnarray}
It becomes exact in the limit of an infinitely heavy (i.e. localized) impurity.
Energy minimization with respect to the variational parameters $\alpha_{\vec{k}}$ gives
 \begin{eqnarray}
\alpha_{\vec{k}}^\MF = - \frac{V_k}{ \Omega^\MF_{\vec{k}}} = - \frac{V_k}{\omega_k + \frac{k^2}{2M} - \frac{\vec{k}}{M} \cdot \l \vec{P} - \vec{P}^\MF_\ph  \r}, 
\label{alpha_k}
 \end{eqnarray}
 where $\vec{P}^\MF_\ph$ is the momentum of the system carried by the phonons. It has to be determined self-consistently from the solution \eqref{alpha_k},
 \begin{eqnarray}
 \vec{P}^\MF_\ph = \int d^d k ~ \vec{k} |\alpha^\MF_{\vec{k}}|^2.
\label{Theta_MF}
 \end{eqnarray}
 The MF character of the wave function (\ref{Psi_MF}) is apparent from the fact that it is a product of wave functions for individual phonon modes. Hence it contains neither entanglement nor correlations between different modes. The only interaction between modes is through the selfconsistency equation (\ref{Theta_MF}).

Properties of the MF solution have been discussed extensively in Refs. \cite{Lee1953,Devreese2013,Shashi2014RF}. Here we reiterate only one important issue 
related to the high energy regularization of the MF energy \cite{Tempere2009}. In $d \geq 2$ dimensions the expression for the MF energy,
\begin{eqnarray}
E_{\rm B}^\MF = \frac{P^2}{2 M} - \frac{\l P_\ph^{\MF} \r^2 }{2M} - \int_{|k|<\Lambda_0} d^d k ~ \frac{V_k^2}{\Omega^\MF_{\vec{k}}},
\label{EBmeanfield}
\end{eqnarray}
is UV divergent as the high momentum cutoff $\Lambda_0$ is sent to infinity. In order to regularize this expression we recall that the physical energy of the impurity is a sum of $E^0_{\rm IB}$ and the polaron binding energy $E_\B$. If we use the leading order Born approximation to express $g_{\rm IB} = 2\pi a_{\rm IB}/m_{\rm red}$, we observe that the MF polaron energy has contributions starting with the second order in $a_{\rm IB}$. Consistency requires that the impurity-condensate interaction energy $E_{\rm IB}^0 = g_{\rm IB} n_0$ is computed to order $a_{\rm IB}^2$. The Lippman-Schwinger equation provides the relation between the microscopic interaction $g_{\rm IB}$, the cutoff $\Lambda_0$ and the physical scattering length $a_{\rm IB}$ \cite{Pethick2008},
\begin{align}
a_\IB  &= \frac{g_\IB}{2 \pi} m_\text{red} - \frac{g_{\IB}^2}{(2\pi)^4} m_\text{red} \int_{|k|<\Lambda_0} d^d k ~ \frac{2 m_\text{red}}{k^2}.
\label{eq:LSE}
\end{align}
To second order in $a_{\rm IB}$ one has
\begin{equation}
E^0_{\rm IB}= \frac{2 \pi a_{\rm IB} n_0}{m_{\rm red}} + \frac{a_\IB^2 n_0}{\pi m_\text{red}} \int_{|k|<\Lambda_0} d^dk ~ \frac{1}{k^2}.
\label{EIB02ndorder}
\end{equation}
Now we recognize that in the physically meaningful impurity energy $E_\text{IMP} = E_\IB^0 + E_\B^\MF$ the UV divergence cancels between $E_{\rm IB}^0$ and
$E_{\rm B}^{\rm MF}$.

Thus separation of the impurity energy into the impurity-condensate interaction $E_\IB^0=n_0 g_\IB$ and the binding energy $E_\B$ is not physically meaningful. A decomposition of the impurity energy in orders of the scattering length $a_\IB$, on the other hand, is well defined,
\begin{equation}
E_{\rm IMP} = \frac{2 \pi a_{\rm IB} n_0}{m_{\rm red}} + E_\p(a_\IB^2) ,
\label{eq:EpDef}
\end{equation}
where the second-order term is referred to as the polaronic energy of the impurity \cite{Tempere2009}. To characterize the polaronic energy, it is convenient to introduce a dimensionless coupling constant,
\begin{equation}
\alpha = 8 \pi n_0 a_\IB^2 \xi,
\end{equation}
highlighting the analogy e.g. to the solid state Fr\"ohlich model.

The main shortcoming of the MF ansatz \eqref{Psi_MF} is that it discards correlations between phonons with different energies and at different momenta. The goal of this paper is to develop a method that allows us to go beyond the MF solution (\ref{Psi_MF}) and include correlations between modes.  In the following we demonstrate how this can be accomplished, and discuss physical consequences of phonon correlations. To simplify the subsequent discussion we perform a unitary transformation that shifts the phonon variables in Eq. \eqref{H_LLP} by the amount corresponding to the MF solution,
\begin{eqnarray}
\hat{V} &=& \exp \left[ \smallint d^d k ~ \alpha_{\vec{k}}^\MF \ad_{\vec{k}} - {\rm h.c.} \right]
\nonumber\\
\tilde{\cal H}_{\rm LLP} &=& \hat{V}^\dagger \H_{\rm LLP} \hat{V}
\nonumber \\
&=& E_\B^\MF + \int_k ~ \Omega^\MF_{\vec{k}} \ad_{\vec{k}} \a_{\vec{k}}  +
\int_{kk'} \frac{A_{\vec{k} \vec{k}'}}{2} :\G_{\vec{k}} \G_{\vec{k}'}:. \qquad
\label{H_fluct}
\end{eqnarray}
Here we used $A_{\vec{k} \vec{k}'} = \frac{\vec{k} \cdot \vec{k}'}{M}$ and $\G_{\vec{k}} := \alpha^\MF_{\vec{k}} ( \a_{\vec{k}} + \ad_{\vec{k}} ) + \ad_{\vec{k}} \a_{\vec{k}}$, $: ... :$ stands for normal-ordering and we introduced the short-hand notation $\int_k = \int_{|k|<\Lambda_0} d^d k$. The absence of terms linear in $\a_{\vec{k}}$ in the last equation reflects the fact that $\alpha_{\vec{k}}^\MF$ correspond to the MF (saddle point) solution. We emphasize that  (\ref{H_fluct}) is an exact representation of the original Fr\"ohlich Hamiltonian, where the operators $\a_{\vec{k}}$ describe quantum fluctuations around the MF polaron.

\section{RG Analysis}
\label{sec:RGana}

\begin{table}[b]
\begin{tabular}{  p{5cm} | p{2.4cm} }
 operator &  scaling ($\Lambda \xi \ll 1$) \\[1ex]
 \hline
 $\a_{\vec{k}}$  & $\Lambda^{-(d+1)/2}$ \\ [0.8ex]
 \hline
 $ \int_{|k|<\Lambda} d^d k ~ d^d k' ~ \frac{\vec{k} \cdot \vec{k}'}{2 M} \alpha^\MF_{\vec{k}} \alpha^\MF_{\vec{k}'} ~ \a_{\vec{k}} \a_{\vec{k}'}$  & $\Lambda^d$ \\[1ex]
  $ \int_{|k|<\Lambda} d^d k ~ d^d k' ~ \frac{\vec{k} \cdot \vec{k}'}{2 M} \alpha^\MF_{\vec{k}} ~ \ad_{\vec{k}'} \a_{\vec{k}'} \a_{\vec{k}}  $  & $\Lambda^{d/2}$ \\[1ex]
  $ \int_{|k|<\Lambda} d^d k ~ d^d k' ~ \frac{\vec{k} \cdot \vec{k}'}{2 M} \ad_{\vec{k}'} \ad_{\vec{k}} \a_{\vec{k}} \a_{\vec{k}'}$ & $\Lambda^0=1$
\end{tabular}
\caption{Dimensional analysis is performed by power-counting of the different terms describing quantum fluctuations around the MF polaron state. We fixed the scaling dimension of $\a_{\vec{k}}$ such that $\int_{|k|<\Lambda} d^dk~ \Omega_{\vec{k}} \ad_{\vec{k}} \a_{\vec{k}} \stackrel{!}{\sim} \Lambda^0$ is scale-invariant.}
\label{tab:dimAn}
\end{table}

In this section we provide the RG solution of the Hamiltonian \eqref{H_fluct}, describing quantum fluctuations on top of the MF polaron state. We begin with a dimensional analysis of different terms in \eqref{H_fluct} in the long wavelength limit, which establishes that only one of the interaction terms is marginal and all others are irrelevant. Then we derive the RG flow equations for parameters of the model, including the expression for the polaron binding energy.

Our approach to the RG treatment of the model (\ref{H_fluct}) is similar to the "poor man's RG" in the context of the Kondo problem. We  use Schrieffer-Wolff type transformation to integrate out high energy phonons in a thin shell in momentum space
near the cutoff, $\Lambda -\delta \Lambda < k < \Lambda$, using $1/\Omega_{\vec{k}}$ as a small parameter ($\Omega_{\vec{k}} = \Omega_{\vec{k}}^\MF$ being the frequency of phonons in the thin momentum-shell). This transformation renormalizes the effective Hamiltonian for the low energy phonons. Iterating this procedure we get a flow of the effective Hamiltonian with the cutoff parameter $\Lambda$
\footnote{For readers not familiar with the RG methods
we note that this procedure is conceptually similar to the Born-Oppenheimer approximation for the problem of interacting ions and electrons in a molecule. Since ions are much heavier (and thus slower) than electrons, one can consider them as static and calculate the ground state of electrons for a given configuration of ions. Then the ground state energy of the electrons becomes an effective potential for the ions. In our case, fast degrees of freedom are phonons in the shell that we are integrating out, and slow degrees of freedom are phonons at lower energy.}.

To analyze whether the system flows to strong or weak coupling in the long wavelength limit $| \vec{k} | \xi \ll 1$ we consider scaling dimensions of different operators in \eqref{H_fluct}. We fix the dimension of $\a_{\vec{k}}$ using the condition that
$
 \int_{|k|<\Lambda} d^dk~ \Omega_{\vec{k}} \ad_{\vec{k}} \a_{\vec{k}}
$ is scale invariant. In the long wavelength limit phonons have linear dispersion $\Omega_{\vec{k}} \propto |\vec{k}|$, which requires scaling of $\a_{\vec{k}}$ as  $\Lambda^{-\frac{d+1}{2}}$. Scaling dimensions of different contributions to the interaction part of the Hamiltonian \eqref{H_fluct} is shown in table \ref{tab:dimAn}. 
We observe that, as the cutoff scale tends to zero, most terms are irrelevant and only the quartic term 
$\int_{kk'} \frac{\vec{k} \cdot \vec{k}'}{2 M} ~ \ad_{\vec{k}'} \ad_{\vec{k}} \a_{\vec{k}} \a_{\vec{k}'}$ is marginal. As we demonstrate below,  this term is marginally irrelevant, i.e. in the process of RG flow the impurity mass $M$ flows to large values. This feature provides justification for doing the RG perturbation expansion with $1/M$ as the interaction parameter. Also, an irrelevance of the interaction under the RG flow physically means that at least slow phonons in the system 
are Gaussian. This provides an insight why the variational correlated Gaussian wavefunctions \cite{Shchadilova2014} are applicable for the Fr\"ohlich Hamiltonian under consideration.

To facilitate the subsequent discussion of the RG procedure we write a generalized form of Eq. \eqref{H_fluct} that allows for the additional terms in the effective Hamiltonian which will be generated in the process of the RG,
\begin{multline}
 \tilde{\cal H}_{\rm RG}(\Lambda) = E_\B+ \int_{|k|<\Lambda} d^d k~ \l  \Omega_{\vec{k}} \ad_{\vec{k}} \a_{\vec{k}} + W_{\vec{k}} ( \ad_{\vec{k}} + \a_{\vec{k}} ) \r \\
+\frac{1}{2} \int_{|k|,|k'|<\Lambda} d^d k ~ d^d k' ~ k_\mu \mathcal{M}_{\mu \nu}^{-1} k_\nu' : \G_{\vec{k}} \G_{\vec{k}'} :.
\label{H_RG} 
 \end{multline}
Note that the interaction is now characterized by a general tensor $\mathcal{M}^{-1}_{\mu\nu}$ \footnote{The indices $\mu=x,y,z,...$ label cartesian coordinates and they are summed over when occurring twice.}, where the anisotropy originates from the total momentum of the polaron $\vec{P} = P \vec{e}_x$, breaking the rotational symmetry of the system. Due to the cylindrical symmetry of the problem, the mass tensor has the form $\mathcal{M}= \text{diag} ( \mathcal{M}_\parallel , \mathcal{M}_\perp, \mathcal{M}_\perp,... )$, and we will find different flows for the longitudinal and the transverse components of the mass tensor. While $\mathcal{M}$ can be interpreted as the (tensor-valued) renormalized mass of the impurity, it should not be confused with the mass of the polaron. The first line of Eq.\eqref{H_RG} describes the diagonal quadratic part of the renormalized phonon Hamiltonian. It is also renormalized compared to the original expression in Eq.\eqref{H_fluct},
\begin{equation}
\Omega_{\vec{k}} = \omega_k + \frac{1}{2} k_\mu \mathcal{M}_{\mu \nu}^{-1} k_\nu - \frac{\vec{k}}{M} \cdot \l \vec{P} - \vec{P}_\ph  \r ,
\label{eq:Ok}
\end{equation}
where the momentum carried by the phonon-cloud, $\vec{P}_\ph$, acquires an RG flow, describing corrections to the MF result $\vec{P}_\B^\MF$. In addition there is a term linear in the phonon operators, weighted by
\begin{equation}
W_{\vec{k}} =  \left[  \l \vec{P}_{\ph} - \vec{P}_{\ph}^\MF \r \cdot \frac{\vec{k}}{M} + \frac{k_\mu  k_\nu}{2} \l \mathcal{M}_{\mu \nu}^{-1} - \frac{\delta_{\mu \nu}}{ M}  \r \right] \alpha_{\vec{k}}^\MF.
\label{eq:defWk}
\end{equation}
By comparing Eq.\eqref{H_RG} to Eq.\eqref{H_fluct} we obtain the initial conditions for the RG, starting at the original UV cutoff $\Lambda_0$ where $ \tilde{\cal H}_{\rm RG}(\Lambda_0) = \tilde{\cal H}_{\rm LLP}$,
\begin{equation}
\mathcal{M}_{\mu \nu}(\Lambda_0) = \delta_{\mu \nu} M, \quad \vec{P}_\ph(\Lambda_0) = \vec{P}_\ph^\MF, \quad E_\B(\Lambda_0)=E_\B^\MF.
\end{equation}

\begin{figure*}[t]
\centering
\epsfig{file=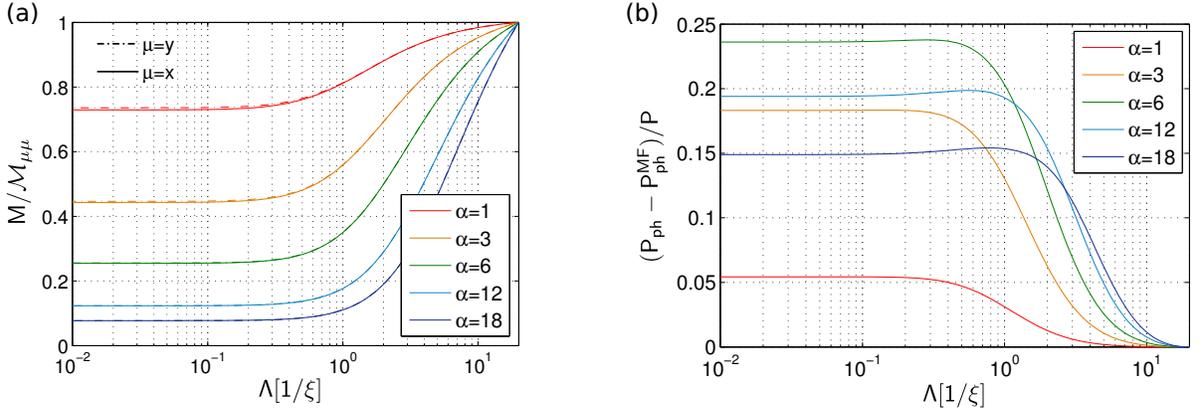, width=0.9\textwidth}
\caption{Typical RG flows of the (inverse) renormalized impurity mass $\mathcal{M}^{-1}$ (a) and the excess phonon momentum $P_\ph-P_\ph^\MF$ along the direction of the system momentum $P$ (b). Results are shown for different coupling strengths $\alpha$ and we used parameters $M/m=0.3$, $P/Mc=0.5$ and $\Lambda_0=20 / \xi$ in $d=3$ dimensions.}
\label{fig:FIG2}
\end{figure*}

We now separate phonons into "fast" ones with momenta $\vec{p}$ and "slow" ones with momenta $\vec{k}$, according to $\Lambda -\delta \Lambda < | \vec{p} | < \Lambda$ and $ | \vec{k} | \leq \Lambda -\delta \Lambda$. Then the Hamiltonian \eqref{H_RG} can be split into 
\begin{eqnarray}
 \tilde{\cal H}_{\rm RG}(\Lambda)  &=& \H_\s + \H_\f +\H_\sf
\nonumber\\
\H_{\rm F} &=& \int_\f d^d p~ \left[  \Omega_{\vec{p}} \ad_{\vec{p}} \a_{\vec{p}} + W_{\vec{p}} ( \ad_{\vec{p}} + \a_{\vec{p}} ) \right]   
\nonumber\\
  {\cal H}_{\rm MIX} &=&  \int_\s d^d k  \int_\f d^d p ~ k_\mu \mathcal{M}_{\mu \nu}^{-1} p_\nu' ~  \G_{\vec{k}} \G_{\vec{p}},
  \label{Hsplit}
\end{eqnarray}
where we use the short-hand notations $\int_\f d^dp = \int_{\vec{p}~ \text{fast}} d^dp$ and $\int_\s d^dk = \int_{\vec{k}~ \text{slow}} d^dk$. The slow-phonon Hamiltonian $\H_\s$ is given by Eq.\eqref{H_RG} except that all integrals only go over slow phonons, $\int_{|\vec{k}|<\Lambda} d^dk \rightarrow \int_\s d^dk$.
In $\H_\f $ we do not have a contribution due to the interaction term since it would be proportional to $\delta \Lambda^2$ and we will consider the limit $\delta \Lambda \rightarrow 0$. We can obtain intuition into the nature of the transformation needed to decouple fast from slow phonons, by observing that for the fast phonons the Hamiltonian \eqref{Hsplit} is similar to a harmonic oscillator in the presence of an external force (recall that $\G_{\vec{p}}$ contains only linear and quadratic terms in $\a_{\vec{p}}^{(\dagger)}$). This external force is determined by the state of slow phonons. Thus it is natural to look for the transformation as a shift operator for the fast phonons, 
\begin{eqnarray}
 \hat{W}_{\rm RG} = \exp \l \int_\f d^3 \vec{p} ~ \left[ \F_{\vec{p}}^\dagger \a_{\vec{p}} - \F_{\vec{p}} \ad_{\vec{p}} \right] \r,
\end{eqnarray} 
with coefficients $\F_{\vec{p}}$ depending on the slow phonons only, i.e. $[\F_{\vec{p}},\a_{\vec{p}}^{(\dagger)}]=0$. One can check that taking 
\begin{widetext}
\begin{multline}
\F_{\vec{p}} =  \frac{1}{\Omega_{\vec{p}}} \left[ 
W_{\vec{p}} + \alpha^\MF_{\vec{p}} p_\mu \mathcal{M}_{\mu \nu}^{-1} 
\int_\s d^d k ~  k_\nu  \G_{\vec{k}}  \right] - 
\frac{1}{\Omega_{\vec{p}}^2} \left[  \alpha^\MF_{\vec{p}} p_\mu \mathcal{M}_{\mu \nu}^{-1}  \int_\s d^d k ~ \Omega_{\vec{k}}^\MF k_\nu \alpha^\MF_{\vec{k}} \l \ad_{\vec{k}} - \a_{\vec{k}} \r + \right. \\ \left. +
\l W_{\vec{p}} + \alpha^\MF_{\vec{p}} p_\mu \mathcal{M}_{\mu \nu}^{-1}
\int_\s d^d k ~  k_\nu  \G_{\vec{k}}  \r  p_\sigma \mathcal{M}_{\sigma \lambda}^{-1} \int_\s d^d k ~  k_\lambda  \G_{\vec{k}} \right]
\label{eq:Fresult}
\end{multline}
eliminates non-diagonal terms in $\a_{\vec{p}}^{(\dagger)}$ up to second order in $1/\Omega_{\vec{p}}$.
After the transformation we find
\begin{eqnarray}
\hat{W}^\dagger_{\rm RG}  &\tilde{{\cal H}}_{\rm RG}&(\Lambda) \hat{W}_{\rm RG} = 
\H_\s + \delta \H_\s 
+ \delta E_0 + \int_\f d^d p ~ \l \Omega_{\vec{p}} + \Delta \hat{\Omega}_{\vec{p}} \r \ad_{\vec{p}} \a_{\vec{p}},
\label{Transformed_H}
\end{eqnarray} 
\end{widetext}
\begin{eqnarray}
\Delta \hat{\Omega}_{\vec{p}} &=& p_\mu \mathcal{M}_{\mu \nu}^{-1}  \int_\s d^d k ~ k_\nu \G_{\vec{k}},
\label{DeltaOmegaSlow}
\\
\delta \H_\s  &=& -   \int_\f d^d p ~ \frac{1}{\Omega_{\vec{p}}} \left[ W_{\vec{p}} +  \alpha_{\vec{p}}^\MF \Delta \hat{\Omega}_{\vec{p}} \right]^2,
\label{dHs}
\end{eqnarray} 
\begin{eqnarray}
\delta E_0 &=& \frac{1}{2} \int_\f d^d p ~ p_\mu \l \mathcal{M}^{-1}_{\mu \nu} - \frac{\delta_{\mu \nu}}{M} \r p_\nu |\alpha_{\vec{p}}^\MF|^2,
 \end{eqnarray} 
which is valid up to corrections of order $1/\Omega_{\vec{p}}^2$ or $\delta \Lambda^2$. The last equation describes a change of the zero-point energy $\delta E_0$ of the impurity in the potential created by the phonons, and it is caused by the RG flow of the impurity mass. To obtain this term we have to carefully treat the normal-ordered term $:\G_{\vec{k}} \G_{\vec{k}'}:$ in Eq.\eqref{H_RG} 
 \footnote{The following relation is helpful to perform normal-ordering, $:\G_{\vec{k}} \G_{\vec{k}'}: = \G_{\vec{k}} \G_{\vec{k}'} - \delta \l \vec{k} - \vec{k}' \r \left[  \G_{\vec{k}} + |\alpha^\MF_{\vec{k}} |^2 \right]$.}. 
 We will show later that this contribution to the polaron binding energy is crucial because it leads to a UV divergence in $d\geq 3$ dimensions.

From the last term in Eq.\eqref{Transformed_H} we observe that the ground state $\ket{\gs}$ of the Hamiltonian is obtained by setting
the occupation number of high energy phonons to zero, $ \bra{\gs} \ad_{\vec{p}} \a_{\vec{p}} \ket{\gs}=0$. Then from Eq.\eqref{dHs} we read off the change in the Hamiltonian for the low energy phonons. From the form of the operator $\Delta \hat{\Omega}_{\vec{p}}$ in Eq.\eqref{DeltaOmegaSlow} one easily shows that the new Hamiltonian $\H_\s+\delta \H_\s$ is of the universal form $\tilde{{\cal H}}_{\rm RG}$, but with renormalized couplings. Thus we derive the following flow equations
for the parameters in $\tilde{{\cal H}}_{\rm RG}  (\Lambda)$,
\begin{eqnarray}
\frac{\partial \mathcal{M}_{\mu \nu}^{-1} }{\partial \Lambda} &=& 2 \mathcal{M}_{\mu \lambda}^{-1}  \int_\f d^{d-1} p ~ \frac{ | \alpha_{\vec{p}}^\MF |^2 }{\Omega_{\vec{p}}} p_\lambda p_\sigma  ~\mathcal{M}_{\sigma \nu}^{-1},
\label{eq:gsFlowM} 
\\
\frac{\partial P_{\ph}^\mu }{\partial \Lambda} &=& - 2 \mathcal{M}_{\mu \nu}^{-1} \int_\f d^{d-1} p ~  \Big[ \l \vec{P}_\ph^\MF - \vec{P}_{\ph} \r \cdot  \vec{p} + 
\nonumber \\
 &&    + \frac{1}{2} p_\sigma \l \delta_{\sigma \lambda} - M \mathcal{M}_{\sigma \lambda}^{-1} \r p_\lambda \Big] \frac{| \alpha_{\vec{p}}^\MF |^2 }{\Omega_{\vec{p}}}  p_\nu. 
\label{eq:gsFlowQ}
\end{eqnarray}
Here we use the notation $\int_\f d^{d-1} p$ for the integral over the $d-1$ dimensional surface defined by momenta of length $|\vec{p}| = \Lambda$. The energy correction to the binding energy of the polaron beyond MF theory, $E_\B = E_\B^\MF + \Delta E_\B^{\rm RG}$, is given by
\begin{multline}
\Delta E_\B^{\rm RG}= - \int_{|k| < \Lambda_0} d^d k \left\{  \frac{1}{\Omega_{\vec{k}} } \left| W_{\vec{k}} \right|^2  +  \right.  \\
\left. + \frac{1}{2}  |\alpha_{\vec{k}}^\MF |^2   k_\mu \left[ \frac{\delta_{\mu \nu}}{M} - \mathcal{M}_{\mu \nu}^{-1}(\Lambda=k)  \right]  k_\nu  \right\}.
\label{Energy_renormalization}
\end{multline}
Note that, in this expression, we evaluated the renormalized impurity mass $\mathcal{M}_{\mu \nu}(\Lambda=k)$ at a value of the running cut-off $\Lambda=k$ given by the integration variable $k=|\vec{k}|$. Similarly, it is implicitly assumed that $\vec{P}_\ph(\vec{k})$ and $\mathcal{M}(\vec{k})$ appearing in the expressions for $W_{\vec{k}}$ and $\Omega_{\vec{k}}$, see Eqs.\eqref{eq:defWk} and \eqref{eq:Ok}, are evaluated at $\Lambda=k$.

FIG.\ref{fig:FIG2} shows typical RG flows of $\mathcal{M}_{\mu\nu}$ and $P_\ph-P_\ph^\MF$. For $\Lambda \lesssim 1 / \xi$ we observe quick convergence of these coupling constants. One can see comparison of the MC and RG calculations \cite{Vlietnick2014} for the polaron binding energy at momentum $P=0$ in FIG.\ref{fig:FIG1}. The agreement is excellent for a broad range of interaction strengths. We will discuss these results further in Sec.\ref{sec:Results}.

\section{Regularizing cutoff dependence}
\label{sec:regularization}

\begin{figure}[t]
\centering
\epsfig{file=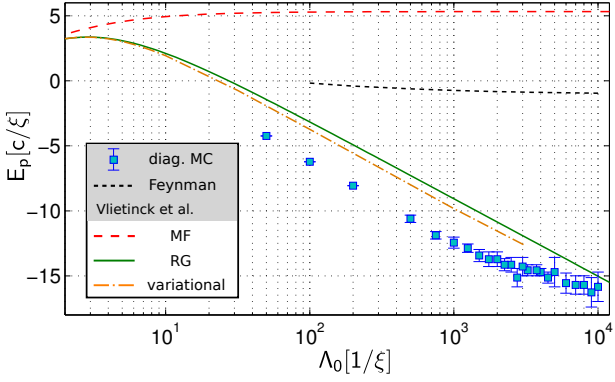, width=0.45\textwidth}
\caption{The polaronic contribution to the ground state energy $E_\p$ (as defined in Eq.\eqref{eq:EpDef}) in $d=3$ dimensions is shown as a function of the UV momentum cutoff $\Lambda_0$ in logarithmic scale. We compare our results (RG - solid, variational \cite{Shchadilova2014} - dash-dotted) to MF theory as well as to predictions by Vlietinck et al. \cite{Vlietnick2014} (diagrammatic MC - squares, Feynman - dashed). The data shows a logarithmic UV divergence of the polaron energy. Parameters are $M/m=0.263158$, $P=0$ and $\alpha=3$.}
\label{fig:FIGlogDiv}
\end{figure}

In a three dimensional system ($d=3$), examination of the binding energy of the polaron $E_{\rm B}$ shows that it has logarithmic UV divergence, $E_\B \sim - \log(\Lambda_0 \xi)$, even after applying the regularization procedure described in Sec.\ref{sec:RevMF}. In FIG.\ref{fig:FIGlogDiv} we show results from the RG (and from our variational calculation introduced in \cite{Shchadilova2014}) and compare them to MC calculations \cite{Vlietnick2014}. While the predicted overall energy scale differs somewhat, a logarithmic divergence can be identified in all three cases \footnote{In their paper \cite{Vlietnick2014}, Vlietinck et al. claim that the polaron energy is UV convergent. When plotting their data on a logarithmic scale as done in our FIG. \ref{fig:FIGlogDiv}, we think that this claim is not justified. Over a range of almost two decades we observe a clearly linear slope. Only for the largest values of the UV cutoff there are some deviations, which are on the order of the statistical errorbars, however.}. 
Similar log-divergences related to vacuum fluctuations are well known from the Lamb-shift in quantum electrodynamics and have been predicted in relativistic polaron models \cite{Volovik2014} as well. In this section we derive the form of the log-divergence analytically using the RG,  and discuss how the binding energy can be regularized. We focus on the zero momentum polaron $P=0$ and the experimentally most relevant case of $d=3$ dimensions, but the extensions to finite polaron momentum $P \neq 0$ or $d \neq 3$ are straightforward.

To check whether the corrections $\Delta E_\B^{\rm RG}$ to the MF polaron binding energy are UV divergent, we need the asymptotic form of the RG flow for the impurity mass $\mathcal{M}(\Lambda=k)$ at large momenta. For $P=0$ the flow equation \eqref{eq:gsFlowM} is separable, and from its exact solution we obtain the asymptotic behavior at the beginning of the RG flow (i.e. for high energies),
\begin{eqnarray}
\frac{M}{\mathcal{M}(\Lambda)} = 1-\frac{32}{3} \frac{n_0 a_\IB^2 m_{\rm red}}{M} \frac{1}{\Lambda} + \mathcal{O}(\Lambda^{-2}),
\label{Masympt}
\end{eqnarray}
where we set $\Lambda_0 =\infty$. Using this result in Eq.\eqref{Energy_renormalization}, simple power-counting shows that the contribution from the zero-point energy of the impurity
\begin{equation}
\Delta E_0^{\rm RG} = - \int_{0}^{\Lambda_0} dk ~ 4 \pi k^2 \frac{|\alpha^\MF_k|^2}{2M} k^2 \l 1- \frac{M}{\mathcal{M}(\Lambda=k)} \r
\label{DeltaE0divergent}
\end{equation}
becomes logarithmically UV divergent (recall that $\alpha^\MF_ k\sim 1/ k^2$ at high momenta $k \gg \xi^{-1}$). The first term $-\int d^3k ~ |W_{\vec{k}}|^2/\Omega_{\vec{k}}$ in Eq.\eqref{Energy_renormalization}, on the other hand, is UV convergent because $W_k \sim k^{-1}$. From Eqs.\eqref{Masympt} and \eqref{DeltaE0divergent} we derive the following form of the UV divergence,
\begin{equation}
\Delta E_{\text{UV}}^{\rm RG} = - \frac{128}{3} \frac{m_{\rm red}}{M^2} n_0^2 a_\IB^4 \log \l \Lambda_0 \xi \r.
\label{eq:EUVRG}
\end{equation}
We find that the slope predicted by this curve is in excellent agreement with the MC data shown in FIG.\ref{fig:FIGlogDiv}.

To regularize this divergence we again need to return to the impurity-condensate interaction energy $E^0_{\rm IB}$. When taking this energy to be $g_{\rm IB} n_0$
we understand that $g_{\rm IB}$ stands for the low energy part of the impurity-boson scattering amplitude, which we need to find from the Lippman-Schwinger equation
$
T= V +VGT
$.
Here $T$ denotes the $T$-matrix, $V$ is the scattering potential and $G$ is the free impurity propagator. In the case of a two particle interaction in vacuum, $G$ is taken as $-\int_{k<\Lambda_0} d^dk \frac{2m_{\rm red}}{k^2} \ket{k} \bra{k}$, and as discussed in Sec.\ref{sec:RevMF} we can
restrict ourselves to the second order in $V$ such that $T= V+VGV$. An important change in the polaronic problem compared to vacuum is that condensate atoms interact with an impurity atom when the latter is dressed by the polaronic cloud. Hence when calculating the impurity-condensate interactions, the $k$-dependence of the impurity mass (due to the RG) should be taken into account. This is achieved by calculating the $k \rightarrow 0$ limit of the scattering amplitude $f_{k \to 0} \stackrel{!}{=} - a_\IB$ from the Lippmann-Schwinger equation with the dressed propagator $G^* = -\int_{k<\Lambda_0} d^dk \frac{2m^*_{\rm red}(k)}{k^2} \ket{k} \bra{k}$, where $(m_{\rm red}^*(k))^{-1} = M^{-1} + \mathcal{M}^{-1}(k)$. A comparison to the MF case Eq.\eqref{eq:LSE}, where the impurity mass does not flow, shows that there is an additional contribution to the impurity-condensate interaction $E^0_{\rm IB}$,
\begin{eqnarray}
\Delta E^0_{\rm IB} =  \frac{4 a_\IB^2 n_0}{m_{\text{red}}} \int^{\Lambda_0} dk ~ \l \frac{m_{\text{red}}^*(k)}{m_{\text{red}}} - 1 \r.
\label{eq:deltaE0IB}
\end{eqnarray}

By using the asymptotic solution \eqref{Masympt} in the last equation, together with the definition of $m_{\text{red}}^*(k)$, it is easy to check that the asymptotic behavior of $\Delta E^0_{\rm IB}$ is the same as $\Delta E_{\text{UV}}^{\rm RG}$, see Eq.\eqref{eq:EUVRG}, but with the opposite sign. Hence the resulting impurity energy is UV convergent. The final expression for the impurity energy, that is now free of all UV divergencies, is given by
\begin{eqnarray}
E_{\rm IMP} = \underbrace{E^\MF_\B}_{\rm Eq.\eqref{EBmeanfield}}  + \underbrace{E_\IB^0}_{\rm Eq.\eqref{EIB02ndorder}} + \underbrace{\Delta E^{\rm RG}_\B}_{\rm Eq.\eqref{Energy_renormalization}} + \underbrace{\Delta E^0_{\rm IB}}_{\rm Eq.\eqref{eq:deltaE0IB}}.
\label{eq:EIMPfinal}
\end{eqnarray}
Let us recapitulate the meaning of all these terms: $E^\MF_\B$ describes the binding energy of the polaron at the MF level; $E_\IB^0$ is the impurity-condensate interaction, and it regularizes the power-law divergence of the MF binding energy; $\Delta E^{\rm RG}_\B$ is the correction to the polaron binding energy due to the RG; $\Delta E^0_{\rm IB}$ originates from the impurity-condensate interaction when taking into account the $k$-dependence of the impurity mass, including a correction that regularizes the log-divergence of the RG binding energy corrections.

\section{Results: polaron energy and effective mass}
\label{sec:Results}

\begin{figure}[t]
\centering
\epsfig{file=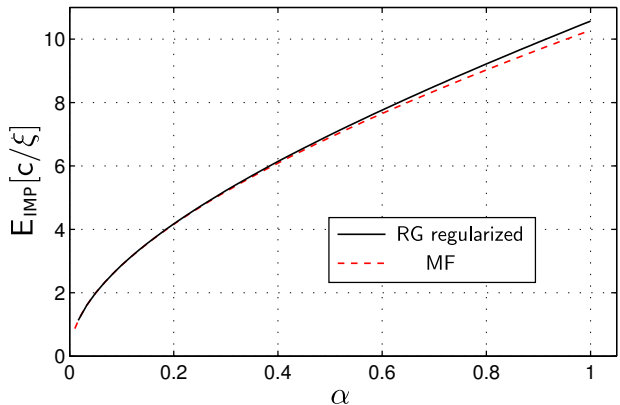, width=0.47\textwidth}
\caption{The impurity energy $E_{\rm IMP}(\alpha)$, which can be measured in a cold atom setup using rf-spectroscopy, is shown as a function of the coupling strength $\alpha$. Our prediction from the RG is given by the solid black line, representing the fully regularized impurity energy from Eq.\eqref{eq:EIMPfinal}. We compare our results to MF theory (dashed). Note that, although MF yields a strict upper variational bound on the binding energy $E_\text{B}$, the MF impurity energy $E_{\rm IMP}$ is below the RG prediction because the impurity-condensate interaction $E_\text{IB}^0$ was treated more accurately in the latter case. We used parameters $M/m=0.26316$, $\Lambda_0=2000 / \xi$, $P=0$ and set the BEC density to $n_0=\xi^{-3}$.}
\label{fig:FIGenergy}
\end{figure}

Now we use the general formalism that we developed in the previous sections to calculate the polaron energy and the effective mass. We find an effective mass that agrees with the MF result for weak coupling and crosses over smoothly to the strong coupling regime. We also compare our analysis to the strong coupling Landau-Pekkar theory.

The fully renormalized impurity energy Eq.\eqref{eq:EIMPfinal} (calculated from the RG) is shown in FIG.\ref{fig:FIGenergy} as a function of the coupling strength $\alpha$. The resulting energy is close to, but slightly above, the MF energy \footnote{{This result might be surprising at first glance, because MF polaron theory relies on a variational principle and thus yields an upper bound for the groundstate energy. However, this bound holds only for the binding energy $E_\B$, defined as the groundstate energy of the Fr\"ohlich Hamiltonian, and not for the entire impurity Hamiltonian including the condensate-impurity interaction.}}. 
This is a consequence of the regularization for the log-divergence introduced in the last section, as can be seen by comparing to curves (MC, RG, variational) where only the power-law divergence was regularized. These curves, on the other hand, are in excellent agreement with each other. We thus conclude that the large deviations observed in FIG.\ref{fig:FIG1} of beyond MF theories (MC, RG, variational) from MF are merely an artifact of the logarithmic UV divergence which was not properly regularized. We note that this also explains -- at least partly -- the unexpectedly large deviations of Feynman's variational approach from the numerically exact MC results, reported in \cite{Vlietnick2014}, since Feynman's model does not capture the log-divergence (see FIG.\ref{fig:FIGlogDiv}). 

Our results have important implications for experiments. The relatively small difference in energy between MF and (properly regularized) RG in FIG.\ref{fig:FIGenergy} demonstrates that a measurement of the impurity energy alone does not allow to discriminate between uncorrelated MF theories and extensions thereof (like RG or MC). Yet such a measurement would still be significant as a consistency check of our regularization scheme. To find smoking gun signatures for beyond MF behavior, other observables are required such as the effective polaron mass, which we discuss next.

\begin{figure}[t]
\centering
\epsfig{file=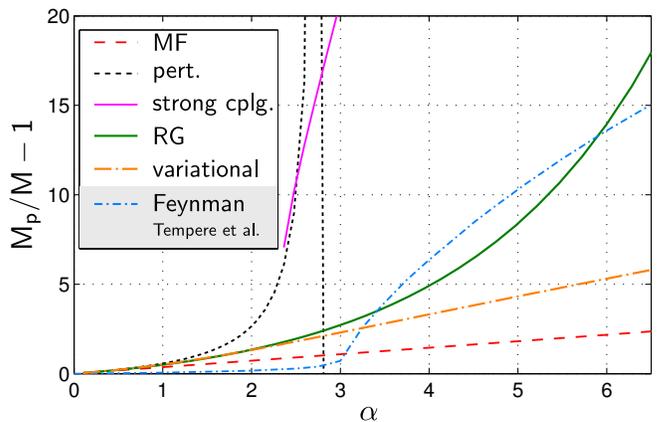, width=0.5\textwidth}
\caption{The polaron mass $M_\p$ (in units of $M$) is shown as a function of the coupling strength $\alpha$. We compare our results (RG) to Gaussian variational \cite{Shchadilova2014} and MF calculations, strong coupling theory \cite{Casteels2011} and Feynman's variational path-integral approach (data taken from Tempere et al.\cite{Tempere2009}). We used parameters $M/m=0.26$, $\Lambda_0=200 / \xi$ and set $P/Mc=0.01$.}
\label{fig:FIGpolaronMass}
\end{figure}

In FIG.\ref{fig:FIGpolaronMass} we show the polaron mass calculated using several different approaches. In the weak coupling limit $\alpha \to 0$ the polaron mass can be calculated perturbatively in $\alpha$, and the lowest-order result is shown in FIG.\ref{fig:FIGpolaronMass}. We observe that in this limit, many approaches follow the same line which asymptotically approaches the perturbative result (as $\alpha \to 0$). One exception is the strong coupling Landau-Pekkar approach, which only yields a self-trapped polaron solution beyond a critical value of $\alpha$ \cite{Casteels2011}. The second exception is Feynman's variational approach, for which we show data by Tempere et al. \cite{Tempere2009} in FIG.\ref{fig:FIGpolaronMass}. The discrepancy observed for small $\alpha$ is surprising since generally Feynman's approach is expected to become exact in the weak coupling limit \cite{Feynman1955,Devreese2013}.

For larger values of $\alpha$, MF theory sets a lower bound for the polaron mass. Naively this is expected, because MF theory does not account for quantum fluctuations due to couplings between phonons of different momenta. These fluctuations require additional correlations to be present in beyond MF wavefunctions, like e.g. in our RG approach, which should lead to an increased polaron mass. Indeed, for intermediate couplings $\alpha \gtrsim 1$ the RG, as well as the variational approach, predict a polaron mass $M_\p > M_\p^\MF$ which is considerably different from the MF result \cite{Shashi2014RF}.

Before proceeding with further discussion of the results, we provide a few specifics on how we calculate the polaron mass. We employ a semi-classical argument to relate the average impurity velocity to the polaron mass $M_\text{p}$ and obtain
\begin{equation}
\frac{M}{M_\text{p}} = 1 - \frac{P_\ph(0)}{P}, \quad P_\ph(0)=\lim_{\Lambda \rightarrow 0}  P_\ph(\Lambda).
\label{eq:MpRG}
\end{equation}
The argument goes as follows. The average polaron velocity is given by $v_\p = P / M_\p$. The average impurity velocity $v_\I$, which by definition coincides with the average polaron velocity $v_\I = v_\p$, can be related to the average impurity momentum $P_\I$ by $v_\I = P_\I/M$. Because the total momentum is conserved, $P = P_\ph + P_\I$, we thus have $P / M_\p = v_\p = v_\I = (P - P_\ph)/M$. Because the total phonon momentum $P_\ph$ in the polaron groundstate is obtained from the RG by solving the RG flow equation in the limit $\Lambda \to 0$, we have $P_\ph = P_\ph(0)$ as defined above, and Eq.\eqref{eq:MpRG} follows. We note that in the MF case this result is exact and can be proven rigorously, see \cite{Shashi2014RF}.

In FIG.\ref{fig:FIGpolaronMass} we present another interesting aspect of our analysis, related to the nature of the cross-over \cite{Gerlach1988,Gerlach1991} from weak to strong coupling polaron regime. While Feynman's variational approach predicts a sharp transition, the RG and variational results show no sign of any discontinuity. Instead they suggest a smooth cross-over from one into the other regime, as expected on general grounds \cite{Gerlach1988,Gerlach1991}. It is possible that the sharp crossover obtained using Feynman's variational approach is an artifact of the limited number of parameters used in the variational action. It would be interesting to consider a more general class of variational actions \cite{GiamarchiEAD}.

\begin{figure}[t]
\centering
\epsfig{file=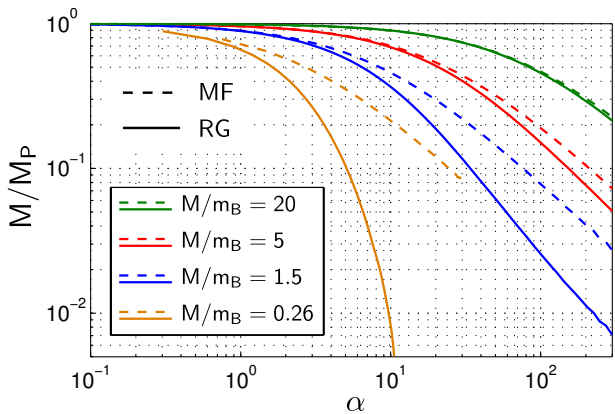, width=0.45\textwidth} $\quad$
\caption{The inverse polaron mass $M/M_\p$ is shown as a function of the coupling strength $\alpha$, for various mass ratios $M/m$. We compare MF (dashed) to RG (solid) results. The parameters are $\Lambda_0=2000 / \xi$ and we set $P/Mc=0.01$ in the calculations.}
\label{fig:FIGpolaronMassLogLog}
\end{figure}

\begin{figure}[t]
\centering
\epsfig{file=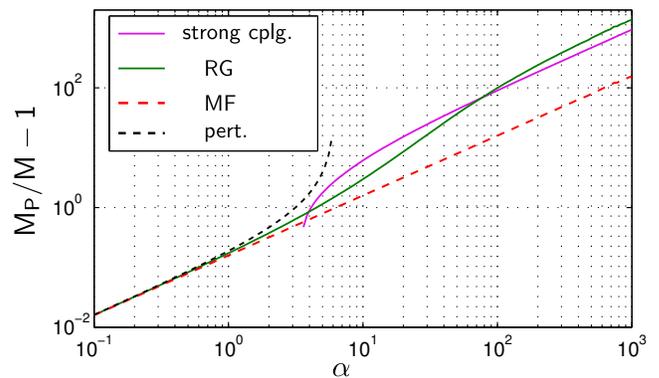, width=0.48\textwidth}
\caption{The polaron mass $M_\p/M$ is shown as a function of the coupling strength for an impurity of mass $M=m$ equal to the boson mass. We compare the asymptotic perturbation and strong coupling theories with MF and RG, which can be formulated for all values of the coupling strength. We used parameters $\Lambda_0=200 / \xi$ and $P/Mc=0.01$.}
\label{fig:FIGpolaronMassCfSC}
\end{figure}

In FIG.\ref{fig:FIGpolaronMass} we calculated the polaron mass in the strongly coupled regime, where $\alpha \gg 1$ and the mass ratio $M/m=0.26$ is small. It is also instructive to see how the system approaches the integrable limit $M \to \infty$ when it becomes exactly solvable \cite{Shashi2014RF}. FIG.\ref{fig:FIGpolaronMassLogLog} shows the (inverse) polaron mass as a function of $\alpha$ for different mass ratios $M/m$. For $M \gg m$, as expected, the corrections from the RG are negligible and MF theory is accurate. When the mass ratio $M/m$ approaches unity, we observe deviations from the MF behavior for couplings above a critical value of $\alpha$ which depends on the mass ratio. Remarkably, for very large values of $\alpha$ the mass predicted by the RG follows the same power-law as the MF solution, with a different prefactor. This can be seen more clearly in FIG.\ref{fig:FIGpolaronMassCfSC}, where the case $M/m=1$ is presented. This behavior can be explained from strong coupling theory. As shown in \cite{Casteels2011} the polaron mass in this regime is proportional to $\alpha$, as is the case for the MF solution. However prefactors entering the weak coupling MF and the strong coupling masses are different.

To make this more precise, we compare the MF, RG and strong coupling polaron masses for $M/m=1$ in FIG.\ref{fig:FIGpolaronMassCfSC}. We observe that the RG smoothly interpolates between the weak coupling MF and the strong coupling regime. While the MF solution is asymptotically recovered for small $\alpha \to 0$ (by construction), this is not strictly true on the strong coupling side. Nevertheless, the observed value of the RG polaron mass in FIG.\ref{fig:FIGpolaronMassCfSC} at large $\alpha$ is closer to the strong coupling result than to the MF theory. 

Now we return to the discussion of the polaron mass for systems with a small mass ratio $M/m<1$. In this case FIG.\ref{fig:FIGpolaronMassLogLog} suggests that there exists a large regime of intermediate coupling, where neither strong coupling nor MF theory can describe the qualitative behavior of the polaron mass. This is demonstrated in FIG.\ref{fig:FIGpolaronMass}, where our RG approach predicts values for the polaron mass midway between MF and strong coupling, for a wide range of couplings. In this intermediate-coupling regime, the impurity is constantly scattered on phonons, leading to strong correlations between them. 

Thus measurements of the polaron mass rather than the binding energy should be a good way to discriminate between different theories describing the Fr\"ohlich polaron at intermediate couplings. Quantum fluctuations manifest themselves in a large increase of the effective mass of polarons, in strong contrast to the predictions of the MF approach based on the wavefunction with uncorrelated phonons. Experimentally, both the quantitative value of the polaron mass, as well as its qualitative dependence on the coupling strength can provide tests of our theory. The mass of the Fermi polaron has successfully been measured using collective oscillations of the atomic cloud \cite{Nascimbene2009}, and we are optimistic that similar experiments can be carried out with Bose polarons in the near future.

\section{Possible experimental realizations}
\label{Supp:ExperimentalConsiderations}
In this Section we discuss conditions under which the Fr\"ohlich Hamiltonian can be used to describe impurities in ultra cold quantum gases. We also present typical experimental parameters and show that the intermediate coupling regime $\alpha \sim 1$ can be reached with current technology. Possible experiments in which the effects predicted in this paper could be observed are also discussed.

To derive the Fr\"ohlich Hamiltonian Eq. \eqref{H_frolich} for an impurity atom immersed in a BEC \cite{Tempere2009,Bruderer2007}, the Bose gas is described in Bogoliubov approximation, valid for weakly interacting BECs. Then the impurity interacts with the elementary excitations of the condensate, which are Bogoliubov phonons. In writing the Fr\"ohlich Hamiltonian to describe these interactions, we included only terms that are linear in the Bogoliubov operators. This implicitly assumes that the condensate depletion $\Delta n$ caused by the impurity is much smaller than the original BEC density, $\Delta n / n_0 \ll 1$, giving rise to the condition \cite{Bruderer2007}
\begin{equation}
 |g_\IB| \ll 4 c  \xi^{2}.
 \label{eq:condBogoFroh}
\end{equation}

To reach the intermediate coupling regime of the Fr\"ohlich model, coupling constants $\alpha$ larger than one $\alpha \gtrsim 1$ are required (for mass ratios $M/m \simeq 1$ of the order of one). This can be achieved by a sufficiently large impurity-boson interaction strength $g_\IB$, which however means that condition \eqref{eq:condBogoFroh} becomes more stringent. Now we discuss under which conditions both $\alpha \gtrsim 1$ and Eq. \eqref{eq:condBogoFroh} can simultaneously be fulfilled. To this end we express both equations in terms of experimentally relevant parameters $a_\text{BB}$ (boson-boson scattering length), $m$ and $M$ which are assumed to be fixed, and we treat the BEC density $n_0$ and the impurity-boson scattering length $a_\IB$ as experimentally tunable parameters. Using the first-order Born approximation result $g_\IB = 2 \pi a_\IB / m_\text{red}$ Eq.\eqref{eq:condBogoFroh} reads
\begin{equation}
\epsilon := 2 \pi^{3/2} \l 1 + \frac{m}{M} \r  a_\IB \sqrt{a_\text{BB} n_0} \stackrel{!}{\ll} 1,
\label{eq:epsExp}
\end{equation}
and similarly the polaronic coupling constant can be expressed as
\begin{equation}
\alpha = 2 \sqrt{2 \pi} \frac{a_\IB^2 \sqrt{n_0}}{\sqrt{a_\text{BB}}}.
\label{eq:alphaExp}
\end{equation}
Both $\alpha$ and $\epsilon$ are proportional to the BEC density $n_0$, but while $\alpha$ scales with $a_\IB^2$, $\epsilon$ is only proportional to $a_\IB$. Thus to approach the strong coupling regime $a_\IB$ has to be chosen sufficiently large, while the BEC density has to be small enough in order to satisfy Eq.\eqref{eq:epsExp}. When setting $\epsilon = 0.3 \ll 1$ and assuming a fixed impurity-boson scattering length $a_\IB$, we find an upper bound for the BEC density,
\begin{multline}
n_0 \leq n_0^\text{max} = 4.9 \times 10^{15} \text{cm}^{-3} \times \l 1 + m/M \r^{-2} \\
\times \l \frac{a_\IB / a_0}{100} \r^{-2} \l \frac{a_\text{BB} / a_0}{100} \r^{-1}, $\qquad$
\label{eq:rhoMax}
\end{multline}
where $a_0$ denotes the Bohr radius. For the same fixed value of $a_\IB$ the coupling constant $\alpha$ takes a maximal value
\begin{equation}
\alpha^{\text{max}} = 0.3\times \frac{\sqrt{2}}{\pi}  \l 1 + m/M \r^{-1} \frac{a_\IB}{a_\text{BB}}
\end{equation}
compatible with condition \eqref{eq:condBogoFroh}.

\begin{table}[t]
 \renewcommand{\arraystretch}{1.4}
\begin{tabular}{c||c|c|c|c|c|c}
\hline
$a_\text{Rb-K}/a_0$ & 284. & 994. & 1704. & 2414. & 3124. & 3834. \\
$\alpha^\text{max}_\text{Rb-K}$ & 0.26 & 0.91 & 1.6 & 2.2 & 2.9 & 3.5 \\
$n_0^{\text{max}} [10^{14} \text{cm}^{-3}]$ &  2.8 & 0.23 & 0.078 & 0.039 & 0.023 & 0.015 \\
\hline 
$a_\text{Rb-Cs}/a_0$ &  ~650. & 1950. & 3250. & 4550. & 5850. & 7150. \\
$\alpha^\text{max}_\text{Rb-Cs}$ &  0.35 & 1.0 & 1.7 & 2.4 & 3.1 & 3.8 \\
$n_0^{\text{max}} [10^{14}  \text{cm}^{-3}]$ & 0.18 & 0.02 & 0.0073 & 0.0037 & 0.0022 & 0.0015 \\
\hline
\end{tabular}
\caption{Experimentally the impurity-boson scattering length $a_\IB$ can be tuned by more than one order of magnitude using a Feshbach-resonance. We consider two mixtures($~^{87}\text{Rb} - ~^{41} \text{K}$, top and $~^{87}\text{Rb} - ~^{133} \text{Cs}$, bottom) and show the maximally allowed BEC density $n_0^\text{max}$ along with the largest achievable coupling constant $\alpha^\text{max}$ compatible with the Fr\"ohlich model, using different values of $a_\IB$.}
\label{tab:alphaMaxNmax}
\end{table}

Before discussing how Feshbach resonances allow to reach the intermediate coupling regime, we estimate values for $\alpha^{\text{max}}$ and $n_0^\text{max}$ for typical background scattering lengths $a_\IB$. Despite the fact that these $a_\IB$ are still rather small, we find that keeping track of condition \eqref{eq:epsExp} is important. To this end we consider two experimentally relevant mixtures, (i) $~^{87}\text{Rb}$ (majority) -$~^{41}\text{K}$ \cite{Catani2008,Catani2012} and (ii) $~^{87}\text{Rb}$ (majority) -$~^{133}\text{Cs}$ \cite{mccarron2011dual,Spethmann2012}. For both cases the boson-boson scattering length is $a_\text{BB}=100 a_0$ \cite{Chin2010,Egorov2013} and typical BEC peak densities realized experimentally are $n_0=1.4 \times 10^{14} \text{cm}^{-3}$ \cite{Catani2008}. In the first case (i) the background impurity-boson scattering length is $a_\text{Rb-K}=284 a_0$ \cite{Chin2010}, yielding $\alpha_\text{Rb-K} = 0.18$ and $\epsilon=0.21<1$. By setting $\epsilon=0.3$ for the same $a_{\text{Rb-K}}$, Eq.\eqref{eq:rhoMax} yields an upper bound for the BEC density $n_0^\text{max}=2.8 \times 10^{14} \text{cm}^{-3}$ above the value of $n_0$, and a maximum coupling constant $\alpha^{\text{max}}_{\text{Rb-K}} = 0.26$. For the second mixture (ii) the background impurity-boson scattering length $a_\text{Rb-Cs}=650 a_0$ \cite{mccarron2011dual} leads to $\alpha_{\text{Rb-Cs}} = 0.96$ but $\epsilon = 0.83 < 1$. Setting $\epsilon=0.3$ for the same value of $a_{\text{Rb-Cs}}$ yields $n_0^\text{max} = 0.18 \times 10^{14} \text{cm}^{-3}$ and $\alpha_{\text{Rb-Cs}}^{\text{max}}=0.35$. We thus note that already for small values of $\alpha \lesssim 1$, Eq.\eqref{eq:epsExp} is \emph{not} automatically fulfilled and has to be kept in mind.

The impurity-boson interactions, i.e. $a_\IB$, can be tuned by the use of an inter-species Feshbach resonance \cite{Chin2010}, available in a number of experimentally relevant mixtures \cite{Park2012,Pilch2009,Ferlaino2006,Ferlaino2006err,Inouye2004,Stan2004,Schuster2012}. In this way, an increase of the impurity-boson scattering length by more than one order of magnitude is realistic. In Table \ref{tab:alphaMaxNmax} we show the maximally achievable coupling constants $\alpha^\text{max}$ for several impurity-boson scattering lengths and imposing the condition $\epsilon < 0.3$. We consider the two mixtures from above ($~^{87}\text{Rb} - ~^{41} \text{K}$ and $~^{87}\text{Rb} - ~^{133} \text{Cs}$), where broad Feshbach resonances are available \cite{Catani2012,Ferlaino2006,Ferlaino2006err,Pilch2009}. We find that coupling constants $\alpha \sim 1$ in the intermediate coupling regime can be realized, which are compatible with the Fr\"ohlich model and respect condition \eqref{eq:condBogoFroh}. The required BEC densities are of the order $n_0 \sim 10^{13} \text{cm}^{-3}$, which should be achievable with current technology. Note that when Eq.\eqref{eq:condBogoFroh} would not be taken into account, couplings as large as $\alpha \sim 100$ would be possible, but then $\epsilon \sim 8 \gg 1$ indicates the importance of the phonon-phonon scatterings neglected in the Fr\"ohlich model.

\section{Conclusion and Outlook}
\label{sec:outlook}
In conclusion, we developed a new method to describe Fr\"ohlich polarons at arbitrary values of the coupling constant. We applied it to analyze the Bose polaron describing impurities immersed in a BEC, which can be realized using mixtures of ultra cold quantum gases. For sufficiently small BEC densities and large interaction strengths, the intermediate coupling regime can be reached with current technology. 

We included quantum fluctuations on top of the MF polaron using an RG approach. Our method predicts polaron energies which are in excellent agreement with numerical diagrammatic MC results and deviate considerably from MF theory. We showed that correlations between phonon modes at different momenta give rise to a logarithmic UV divergence of the polaron energy. Our RG analysis allowed us to understand the origin of this divergence and introduce a corresponding regularization scheme. Furthermore we applied our method to calculate the effective mass of the polaron, which shows a smooth cross-over from weak to strong coupling regime. For impurities lighter than the bosons, we identified an extended regime of intermediate coupling strengths. We point out quantitative agreement between our RG analysis and results based on variational correlated Gaussian wavefunctions in Ref. \cite{Shchadilova2014}. In the future we will also explore polaron physics at larger momenta, where in particular the transition from subsonic- to supersonic regime is poorly understood.

Our predictions can be tested in current experiments with ultra cold atomic mixtures \cite{Palzer2009,Catani2012,Fukuhara2013,Spethmann2012,Scelle2013}. In particular we suggest to measure the effective mass of the polaron, which can provide smoking gun signatures for beyond MF behavior. The effective mass can experimentally be measured from the interaction-induced frequency shift of the impurity in a harmonic trapping potential \cite{Nascimbene2009}. Alternatively polaron properties can be measured using radio-frequency (rf) spectroscopy \cite{Schirotzek2009}, see inset of FIG.\ref{fig:FIG1}. The rf-spectrum directly yields the energy of the impurity immersed in the BEC, and a momentum-resolved variant allows a direct measurement of the polaron dispersion relation. 

Experiments with ultra cold atoms are well suited to investigate non-equilibrium dynamics of polarons \cite{Palzer2009,Catani2012,Fukuhara2013}. Our RG method can be generalized to describe dynamical polaron problems. This allows us, for instance, to calculate the full rf-spectrum of impurities immersed in a BEC, for which considerable deviations are found from MF results \cite{Shashi2014RF} as will be shown in a forthcoming publication. Moreover the problem of polaron formation can be investigated, as well as dynamics of impurities close to the subsonic-supersonic transition. These questions become particularly interesting in the presence of a lattice potential, where MF theory predicts a phase-transition at the edges of the Brillouin zone \cite{Grusdt2013BOprep}.

\section*{Acknowledgements}
We acknowledge useful discussions with I. Bloch, S. Das Sarma, M. Fleischhauer, T. Giamarchi, S. Gopalakrishnan, W. Hofstetter, M. Oberthaler, D. Pekker, A. Polkovnikov, L. Pollet, N. Prokof'ev, R. Schmidt,  V. Stojanovic, L. Tarruell, N. Trivedi, A. Widera and M. Zwierlein. We are indebted to Aditya Shashi and Dmitry Abanin for invaluable input in the initial phase of the project. F.G. is a recipient of a fellowship through the Excellence Initiative (DFG/GSC 266) and is grateful for financial support from the "Marion K\"oser Stiftung". Y.E.S. and A.N.R. thank the Dynasty foundation for financial support. The authors acknowledge support from the NSF grant  DMR-1308435, Harvard-MIT CUA, AFOSR New Quantum Phases of Matter MURI, the ARO-MURI on Atomtronics, ARO MURI Quism program.

\end{document}